\documentclass[twocolumn,showpacs,preprintnumbers,amsmath,amssymb,nofootinbib,superscriptaddress]{revtex4-1}

\usepackage[utf8]{inputenc}
\usepackage{graphicx}
\usepackage{amssymb}
\usepackage{textcomp}
\usepackage{amsmath}
\usepackage{tabularx}
\usepackage{bm}
\usepackage{times}
\usepackage{color}
\usepackage{slashed}
\usepackage{multirow}
\usepackage{verbatim}
\usepackage{cancel}
\usepackage{subfigure}

\usepackage[colorlinks=true, pdfstartview=FitV, linkcolor=blue, citecolor=blue, urlcolor=blue]{hyperref}
\allowdisplaybreaks[4]

\def\lsim{\mathrel{\raise.3ex\hbox{$<$\kern-.75em\lower1ex\hbox{$\sim$}}}}
\def\gsim{\mathrel{\raise.3ex\hbox{$>$\kern-.75em\lower1ex\hbox{$\sim$}}}}

\definecolor{orange}{rgb}{1,0.5,0}

\begin{document}

\title{Connecting Primordial Black Hole to boosted sub-GeV Dark Matter through neutrino}

\author{Wei Chao}
\email{chaowei@bnu.edu.cn}
\affiliation{
Center for Advanced Quantum Studies, Department of Physics,
Beijing Normal University, Beijing 100875, China
}
\author{Tong Li}
\email{litong@nankai.edu.cn}
\affiliation{
School of Physics, Nankai University, Tianjin 300071, China
}
\author{Jiajun Liao}
\email{liaojiajun@mail.sysu.edu.cn}
\affiliation{
School of Physics, Sun Yat-Sen University, Guangzhou 510275, China
}

\begin{abstract}
The explorations of alternative dark matter (DM) candidates beyond WIMP motivated primordial black holes (PBHs) or sub-GeV DM particle in the Milky Way. Neutrinos from PBH evaporation at the present times play as a novel medium boosting sub-GeV DM and leaving signatures in the terrestrial experiments. We explore the boosted DM by the neutrino flux from PBH evaporation (PBH$\nu$BDM) so as to connect the macroscopic PBHs to sub-GeV DM particle. We consider this PBH$\nu$BDM scenario to interpret the XENON1T keV excess. The projected bounds on the sub-GeV DM-electron scattering cross section and the fraction of DM composed of PBHs $f_{\rm PBH}$ are imposed for future experiments.
\end{abstract}

\maketitle

\section{Introduction}
\label{sec:Intro}

Astrophysical evidence suggests that dark matter (DM) comprises 84\% of the matter in the Universe.
The constitution and the characteristics of DM still remain unknown. One usually studied candidate of thermal DM is a weakly interacting massive particle (WIMP) with masses of the order of TeV and weak scale interactions. Due to null conclusive evidence of WIMP at DM direct detection (DD) experiments for a long time~\cite{Schumann:2019eaa}, however, both theoretical and experimental programs have moved the focus to the paradigms beyond the WIMP. The explorations of alternative DM candidates motivated macroscopic objects such as primordial black holes (PBHs)~\cite{Zeldovich:1967lct} or a hypothetical light particle with sub-GeV mass~\cite{Griest:1990kh,Essig:2011nj,Essig:2017kqs}.

It is well known that PBHs with the mass $\sim 10^{15}$ g would evaporate due to Hawking radiation~\cite{Hawking:1974rv} and cannot provide all the observed abundance of DM~\cite{Barrau:2003xp}. The emitted particles such as gamma-rays and $e^\pm$ in the evaporation process are subject to a variety of constraints~\cite{Carr:2020gox}. The neutrinos emitted from PBHs were also studied to a large extent. The MeV neutrinos from PBH evaporation emerge above the diffuse supernova neutrino background (DSNB) and the atmospheric neutrinos~\cite{Calabrese:2021zfq}. It is thus intriguing to explore the possibility of confining light PBHs with current and future terrestrial facilities. The current bounds on the fraction of DM in the form of PBHs were recently obtained from Super-Kamiokande~\cite{Dasgupta:2019cae} and the expected sensitivities were projected by coherent elastic neutrino-nucleus scattering (CE$\nu$NS)~\cite{Calabrese:2021zfq}, JUNO~\cite{Wang:2020uvi}, DUNE and THEIA~\cite{DeRomeri:2021xgy}.

On the other hand, the DM in the Milky Way (MW) halo can be in part composed of a non-thermal light DM which gets boosted to a semi-relativistic velocity by cosmic rays (CRs). The CR boosted DM (BDM, denoted by $\chi$) was proposed to induce novel signatures in the terrestrial experiments~\cite{Cappiello:2018hsu,Yin:2018yjn,Bringmann:2018cvk,Ema:2018bih} and was recently assumed to interpret~\cite{Kannike:2020agf} the XENON1T excess of keV electron recoil events~\cite{XENON:2020rca}. Neutrinos can play as a novel medium boosting DM since they both have weak interactions with other SM particles. Neutrinos and electrons may also share a common strength coupled to DM in some leptophilic neutrino-portal DM model~\cite{Fox:2008kb}. Recently, there proposed boosted DM from solar neutrino~\cite{Jho:2021rmn} or diffuse supernova neutrinos~\cite{Das:2021lcr}.

In this work we explore the boosted DM by the neutrino flux from PBH evaporation (denoted by PBH$\nu$BDM below) so as to connect the macroscopic PBHs to sub-GeV DM particle. We assume the DM component is composed of PBH DM and particle DM in the Milky Way (MW) halo. The PBHs with mass $\sim 10^{14}-10^{16}$ g evaporate MeV scale neutrino flux in the present Universe. The PBH neutrinos scatter with the DM in the MW halo and boost the light components to (semi-)relativistic velocities. Then PBH$\nu$BDM travels through the Earth and scatters with the electrons in the underground detector targets of the low-energy DD experiments. For illustration we assume the $\nu$-DM scattering cross section $\sigma_{\nu\chi}$ in the MW the same as the DM-electron scattering cross section $\sigma_{\chi e}$ in the DD detectors. We consider the PBH$\nu$BDM scenario to interpret the XENON1T keV excess. We then investigate the projected bounds on the cross section $\sigma_{\chi e}$ and the fraction of DM composed of PBHs $f_{\rm PBH}$ in future DD experiments.

This paper is organized as follows. In Sec.~\ref{sec:PBH} we evaluate the neutrino spectra from PBH evaporation. The DM-neutrino scattering in the MW halo and the BDM flux from PBH neutrinos are then calculated in Sec.~\ref{sec:nuboost}. We also compute the DM-electron scattering in the terrestrial facilities and place the current constraint and projected sensitivities. Our conclusions are drawn in Sec.~\ref{sec:Con}.

\section{The neutrino spectra from PBH evaporation}
\label{sec:PBH}

The PBHs smaller than about $10^{15}$ g are light enough to have quantum properties and thermally radiate with a temperature~\cite{Hawking:1975vcx,Page:1976df,Page:1977um,MacGibbon:1990zk,MacGibbon:1991tj}
\begin{eqnarray}
T_{\rm PBH}= {\hbar c^3\over 8\pi G M_{\rm PBH} k_{\rm B}}\approx 10^{-7}\Big({M_{\rm PBH}\over M_\odot}\Big)^{-1} {\rm K}\;,
\end{eqnarray}
where $G$ denotes the Newtonian constant of gravitation, $M_{\rm PBH}$ is the PBH mass and $k_{\rm B}$ is the Boltzmann constant.
We first adopt the public code BlackHawk~\cite{Arbey:2019mbc} to calculate the differential number of neutrinos per unit time emitted by PBHs~\cite{Hawking:1971ei,Page:1976df,Page:1976ki}
\begin{eqnarray}
{d^2N_\nu\over dE_\nu dt}={1\over 2\pi}{\Gamma_\nu(E_\nu,M_{\rm PBH})\over {\rm exp}(E_\nu/k_{\rm B}T_{\rm PBH})+1}\;,
\end{eqnarray}
where $E_\nu$ is the emitted neutrino energy and $\Gamma_\nu$ is the function of greybody factor which encodes the probability of an
elementary spin-1/2 neutrino to escape the PBH gravitational well. We ignore the spin of PBH for simplicity. There are two contributions to the above neutrino spectra generated by the Hawking radiation of PBHs. One of them is the primary contribution consisting of neutrinos directly emitted in the evaporation. The other one is the secondary contribution which origins from the hadronization and the subsequent decay of the primary particles.

\begin{figure}[htb!]
\begin{center}
\includegraphics[scale=1,width=0.8\linewidth]{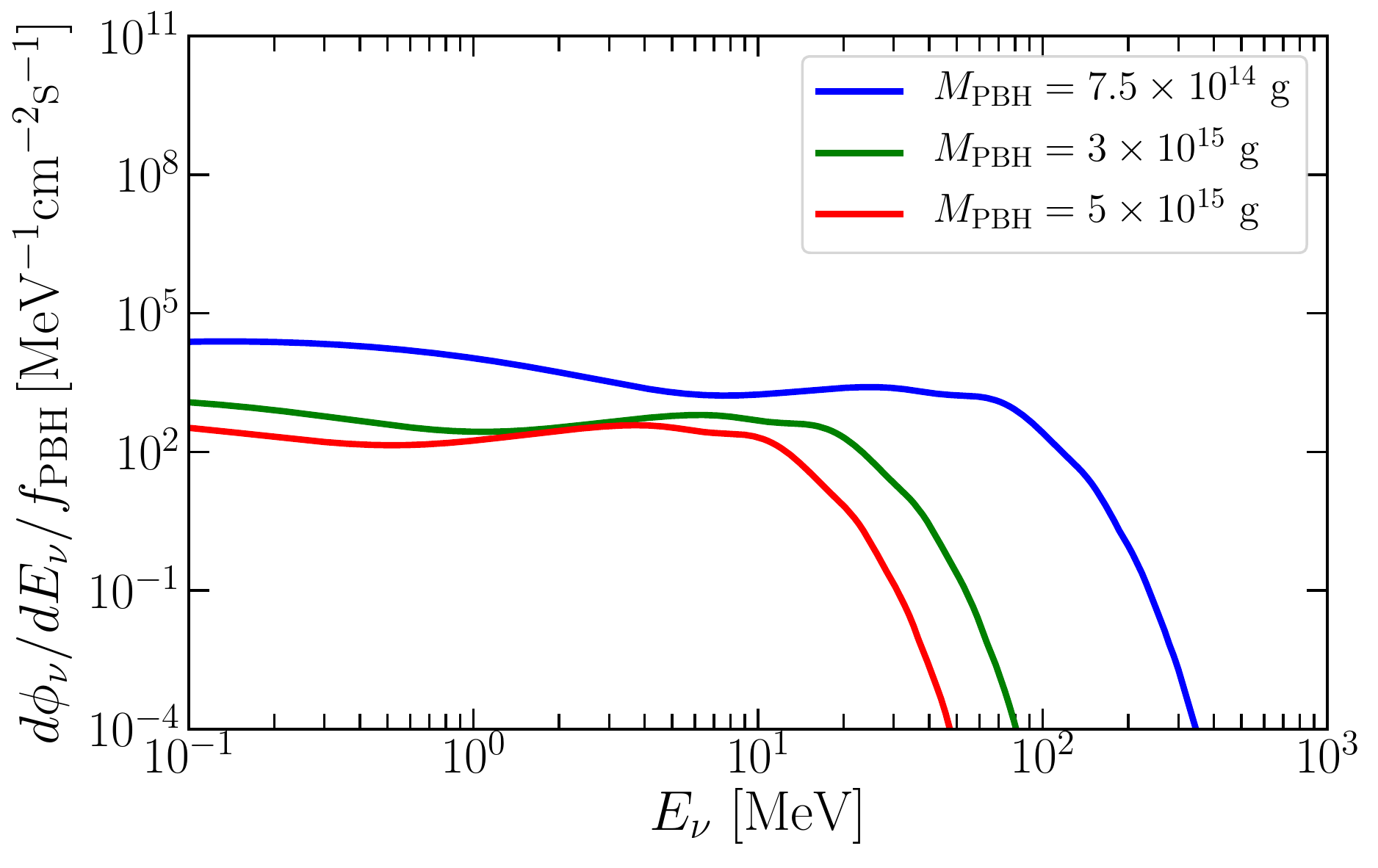}
\end{center}
\caption{The differential neutrino flux from PBHs $f_{\rm PBH}^{-1}d\phi_\nu/dE_\nu$ for three benchmark values of $M_{\rm PBH}=7.5\times 10^{14}$ g, $3\times 10^{15}$ g, $5\times 10^{15}$ g.
}
\label{fig:PBHnuflux}
\end{figure}

For the neutrino flux from PBH evaporation, we consider the contributions of both the PBHs in galactic halo and
the extragalactic PBHs~\cite{Wang:2020uvi,Calabrese:2021zfq,DeRomeri:2021xgy}
\begin{eqnarray}
{d\phi_\nu\over dE_\nu}={d\phi_\nu^{\rm MW}\over dE_\nu}+{d\phi_\nu^{\rm EG}\over dE_\nu}\;.
\label{eq:PBHnu}
\end{eqnarray}
The differential galactic neutrino flux is given by
\begin{eqnarray}
{d\phi_\nu^{\rm MW}\over dE_\nu}={f_{\rm PBH}\over M_{\rm PBH}} {d^2 N_\nu\over dE_\nu dt} \int {d\Omega\over 4\pi} \int dl \rho_{\rm MW}[r(l,\psi)]\;,
\label{eq:MWnuflux}
\end{eqnarray}
where $f_{\rm PBH}$ denotes the fraction of DM composed of PBHs, $\Omega$ is the considered solid angle, $\rho_{\rm MW}[r(l,\psi)]$ is the DM density of the Milky Way (MW) halo, $r(l,\psi)=\sqrt{r_\odot^2+l^2-2lr_\odot \cos\psi}$ is the galactocentric distance with $r_\odot$ the solar distance from the galactic center, $l$ the line-of-sight distance to the PBH and $\psi$ the angle between these two directions. For illustration, we employ the generalized Navarro-Frenk-White (NFW) DM profile~\cite{Navarro:1996gj}
\begin{eqnarray}
\rho_{\rm MW}(r)=\rho_\odot \Big({r\over r_\odot}\Big)^{-\gamma} \Big({1+r_\odot/r_s\over 1+r/r_s}\Big)^{3-\gamma}\;,
\end{eqnarray}
where $\rho_\odot=0.4~{\rm GeV}/{\rm cm}^3$ is the local DM density, $r_\odot=8.5$ kpc, $r_s=20$ kpc is the radius of the galactic diffusion disk, and we fix the inner slope of the NFW halo profile to $\gamma=1$. For the subdominant extragalactic contribution, the corresponding differential neutrino flux over the full sky is
\begin{eqnarray}
{d\phi_\nu^{\rm EG}\over dE_\nu}={f_{\rm PBH}\rho_{\rm DM}\over M_{\rm PBH}}\int^{t_{\rm max}}_{t_{\rm min}} dt [1+z(t)]{d^2N_\nu\over dE_0 dt}\Big|_{E_0=[1+z(t)]E_\nu}\;,
\label{eq:EGnuflux}
\end{eqnarray}
where $\rho_{\rm DM}=2.35\times 10^{-30}~{\rm g}/{\rm cm}^3$ is the average DM density of the Universe at the
present epoch determined by Planck~\cite{Planck:2018vyg}, $E_0$ denotes the neutrino energy at the source and is related to the energy $E_\nu$ in the observer's frame by the redshift $z(t)$. For the integral limits, we fix $t_{\rm min}=10^{11}$ s being close to the era of matter-radiation equality~\cite{Wang:2020uvi} and $t_{\rm max}=\tau_0$ with $\tau_0$ being the age of Universe. In Fig.~\ref{fig:PBHnuflux} we show the differential neutrino flux from PBHs $f_{\rm PBH}^{-1}d\phi_\nu/dE_\nu$ for three benchmark values of $M_{\rm PBH}$. As seen in Eqs.~(\ref{eq:MWnuflux}) and (\ref{eq:EGnuflux}), smaller PBHs exhibit harder spectra of the evaporated neutrinos. The PBH neutrinos have maximal energies of order $E_\nu\sim \mathcal{O}(100)$ MeV for $M_{\rm PBH}\lesssim 10^{15}$ g.

\section{The boosted Dark Matter by PBH neutrinos}
\label{sec:nuboost}

\subsection{DM-neutrino scattering in the MW halo}

\begin{figure}[htb!]
\begin{center}
\includegraphics[scale=1,width=0.8\linewidth]{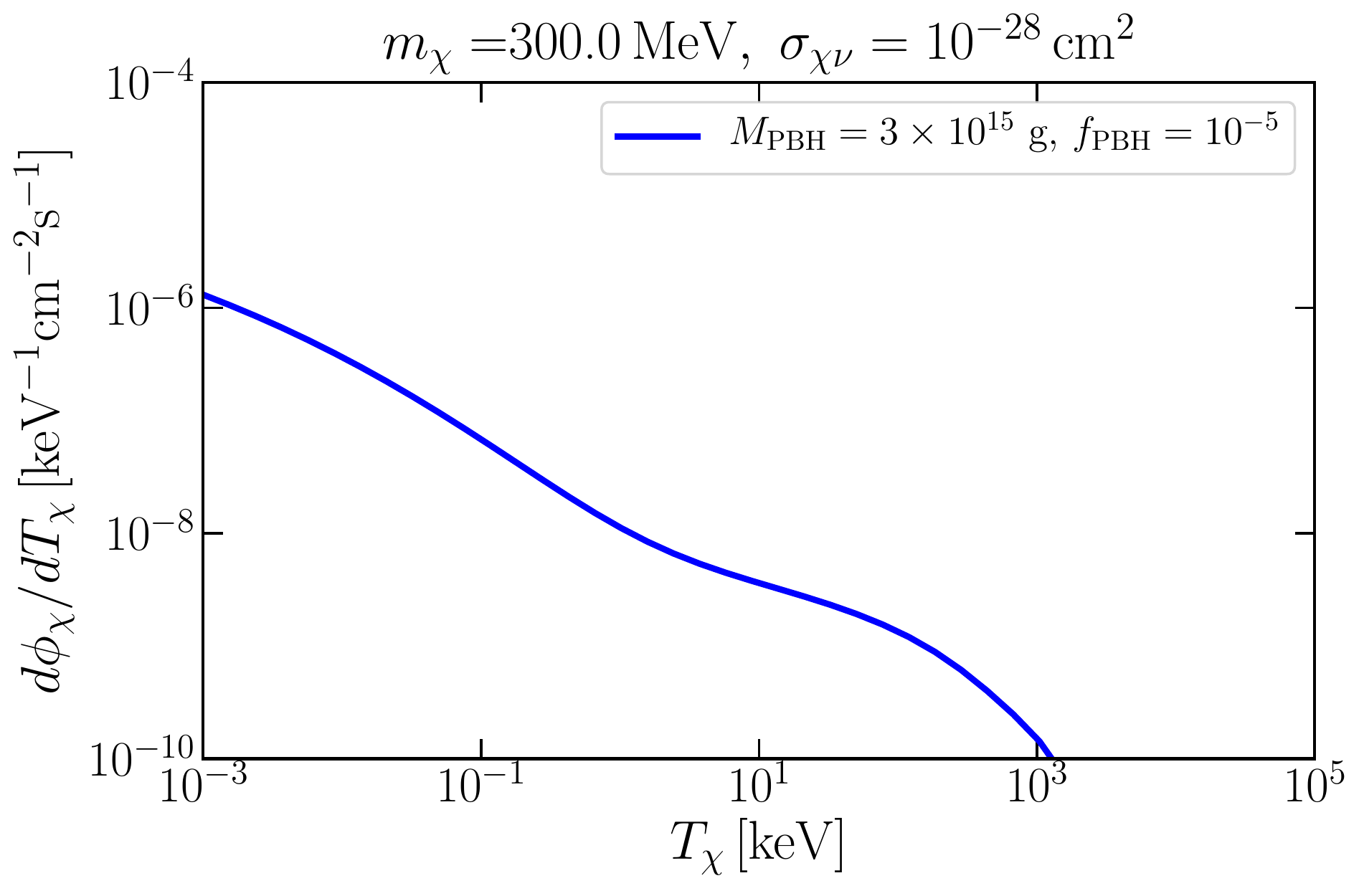}
\end{center}
\caption{The PBH$\nu$BDM flux. For illustration we assume $M_{\rm PBH}=3\times 10^{15}$ g, $f_{\rm PBH}=10^{-5}$, $m_\chi=300$ MeV, and $\sigma_{\chi \nu}=10^{-28}~{\rm cm}^2$.
}
\label{fig:BDMflux}
\end{figure}

The PBH neutrinos can scatter off the particle DM $\chi$ in the MW halo and boost the DM to a much higher velocity. The total upscattered flux of the DM integrated over the whole MW halo is given by~\cite{Das:2021lcr}
\begin{eqnarray}
{d\phi_\chi\over dE_\nu}&=&\int {d\Omega\over 4\pi} \int dl \rho_{\rm MW}[r(l,\psi)] {\sigma_{\nu\chi}\over m_\chi} {d\phi_\nu\over dE_\nu} \nonumber \\
&=&D_{\rm halo}  {\sigma_{\nu\chi}\over m_\chi} {d\phi_\nu\over dE_\nu} \;,
\label{eq:DMflux}
\end{eqnarray}
where $D_{\rm halo}=2.22\times 10^{25}~{\rm MeV}~{\rm cm}^{-2}$, and $\sigma_{\nu\chi}$ denotes the DM-neutrino scattering cross section. In a single scattering of particle DM $\chi$ and neutrino, the neutrino with energy $E_\nu$ can transfer the energy $T_\chi$ to DM $\chi$
\begin{eqnarray}
T_\chi=T_\chi^{\rm max}{1-\cos\theta_{\rm cm}\over 2}\;,~~~T_\chi^{\rm max}={E_\nu^2\over E_\nu+m_\chi/2}\;,
\end{eqnarray}
where we ignore the neutrino mass and $\theta_{\rm cm}$ is the scattering angle in the center-of-mass frame.
Then, in terms of the PBH neutrino flux in Eq.~(\ref{eq:PBHnu}) and the upscattered flux of the DM in Eq.~(\ref{eq:DMflux}), the boosted DM flux can be given by
\begin{eqnarray}
{d\phi_\chi\over dT_\chi}=\int dE_\nu {d\phi_\chi\over dE_\nu} {1\over T_\chi^{\rm max}(E_\nu)} \Theta[T_\chi^{\rm max}(E_\nu)-T_\chi]\;,
\label{eq:BDMflux}
\end{eqnarray}
where $\Theta[T_\chi^{\rm max}(E_\nu)-T_\chi]$ is the Heaviside step function. Fig.~\ref{fig:BDMflux} displays the PBH$\nu$BDM flux in the unit of ${\rm keV}^{-1}~{\rm cm}^{-2}~{\rm s}^{-1}$, with $M_{\rm PBH}=3\times 10^{15}$ g, $f_{\rm PBH}=10^{-5}$, $m_\chi=300$ MeV, and $\sigma_{\chi \nu}=10^{-28}~{\rm cm}^2$ for illustration. This value of the PBHs fraction $f_{\rm PBH}$ is allowed by the evaporation constraints~\cite{Carr:2020gox}.

\subsection{DM-electron scattering in the terrestrial facilities}

The boosted DM $\chi$ can travel a distance underground and scatter with the electrons in the detector of the terrestrial facilities.
Given the PBH$\nu$BDM flux in Eq.~(\ref{eq:BDMflux}), the electron recoil rate can be obtained~\cite{Das:2021lcr}
\begin{eqnarray}
{dR\over dT_e}={Z_\mathcal{N}\over m_\mathcal{N}}\int dT_\chi {d\phi_\chi\over dT_\chi} {1\over T_e^{\rm max}(T_\chi)} \sigma_{\chi e} \;,
\end{eqnarray}
where $Z_\mathcal{N}$($\sim 40$ for Xenon~\cite{Fornal:2020npv}) and $m_\mathcal{N}$ are the effective atomic number and mass of a nucleus $\mathcal{N}$. The maximally allowed recoil energy is
\begin{eqnarray}
T_e^{\rm max}={T_\chi^2+2m_\chi T_\chi\over T_\chi+(m_\chi+m_e)^2/(2m_e)}\;.
\end{eqnarray}
In the following we assume $\sigma_{\nu\chi}=\sigma_{\chi e}$ in a leptophilic neutrino-portal DM model as mentioned in the Introduction.
Following Ref.~\cite{Das:2021lcr}, we adopt a Gaussian detector response function for the electron recoil spectrum
\begin{eqnarray}
\sigma(E)=a\sqrt{E}+bE\;,
\end{eqnarray}
where $a=0.31~\sqrt{\rm keV}$ and $b=0.0037$~\cite{XENON:2020rca}.

\begin{figure}[htb!]
\begin{center}
\includegraphics[scale=1,width=0.8\linewidth]{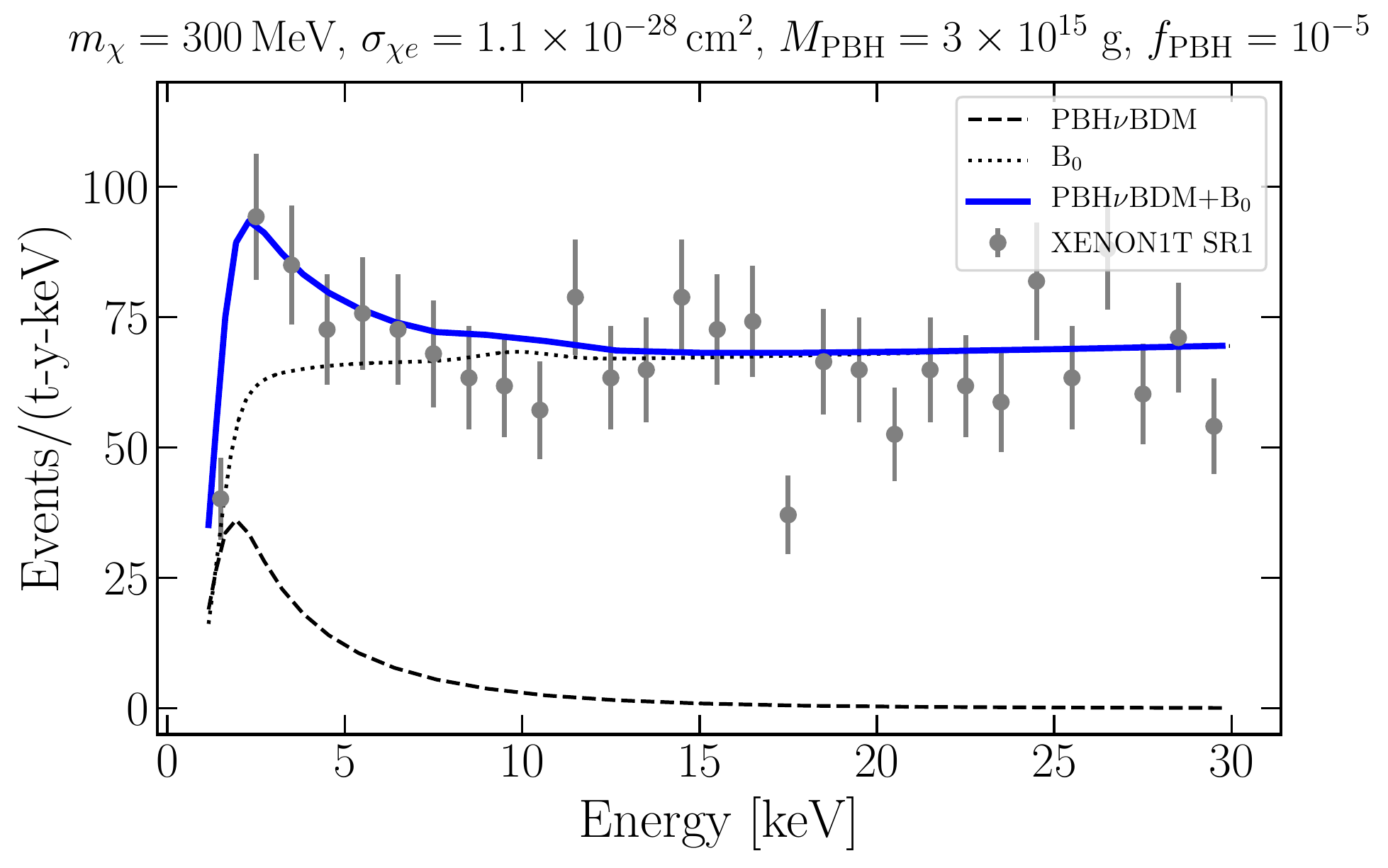}\\
\includegraphics[scale=1,width=0.8\linewidth]{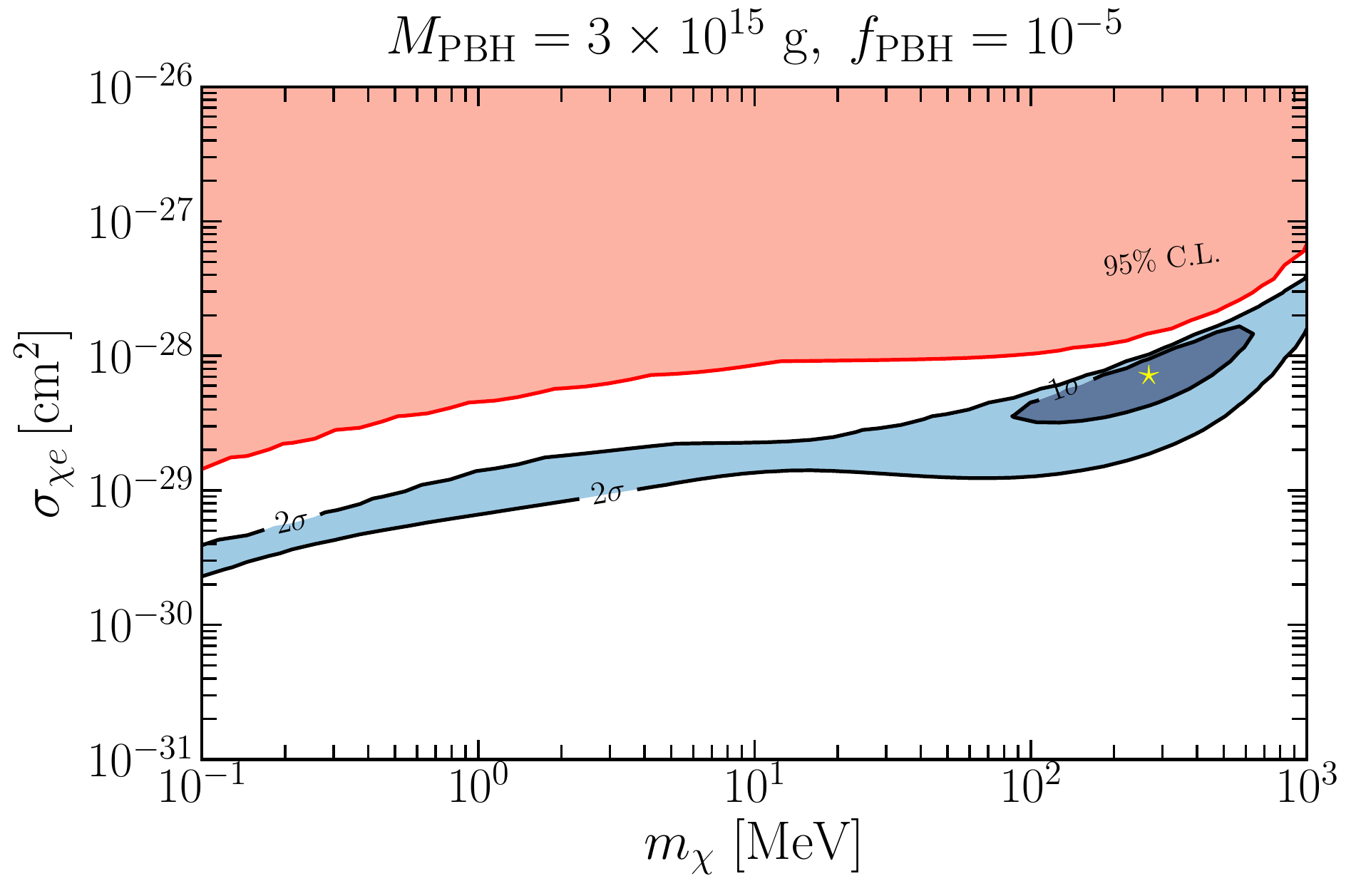}
\end{center}
\caption{Top: The electron recoil spectrum in XENON1T induced by PBH$\nu$BDM without (black dashed) and with (blue solid) the background model spectrum $B_0$ (black dotted). We take $M_{\rm PBH}=3\times 10^{15}$ g, $f_{\rm PBH}=10^{-5}$, $m_\chi=300$ MeV, and $\sigma_{\chi e}=1.1\times 10^{-28}~{\rm cm}^2$.  Bottom: The 1-$\sigma$ (blue) and 2-$\sigma$ (light blue) regions in the $m_\chi-\sigma_{\chi e}$ plane fitted by the XENON1T excess. We assume $M_{\rm PBH}=3\times 10^{15}$ g, $f_{\rm PBH}=10^{-5}$. The best-fit point is marked by a yellow star. The red region is excluded at 95\% confidence level.
}
\label{fig:xenon1t}
\end{figure}

The above details have been implemented in Ref.~\cite{SnBDM} for diffuse supernova neutrinos and the $3\sigma$ excess of electronic recoil events with the energy around 2-3 keV in XENON1T experiment can be explained by supernova neutrinos boosted DM~\cite{Das:2021lcr}. Here we can also explain the keV excess in XENON1T data with regard to PBH$\nu$BDM scenario.
As shown in the top panel of Fig.~\ref{fig:xenon1t}, assuming $M_{\rm PBH}=3\times 10^{15}$ g, $f_{\rm PBH}=10^{-5}$, $m_\chi=300$ MeV, and $\sigma_{\chi e}=1.1\times 10^{-28}~{\rm cm}^2$, one can see that the PBH$\nu$BDM scenario yields a good fit to the XENON1T data taking into account the background model $B_0$~\cite{XENON:2020rca}. In the bottom panel of Fig.~\ref{fig:xenon1t}, for $M_{\rm PBH}=3\times 10^{15}$ g and $f_{\rm PBH}=10^{-5}$, we show the 1-$\sigma$ (blue) and 2-$\sigma$ (light blue) regions in the $m_\chi-\sigma_{\chi e}$ plane preferred by the XENON1T excess. The best-fit point is marked by a yellow star for $m_\chi\sim 300$ MeV and $\sigma_{\chi e}\sim 10^{-28}~{\rm cm}^2$. Following Ref.~\cite{Fermi-LAT:2015qzw,Das:2021lcr}, we consider the events only predicted by the PBH$\nu$BDM scenario and take into account the bins in which they are more than the observed ones. One can then obtain the 95\% confidence level (CL) exclusion region of $\sigma_{\chi e}$ as displayed by the red region.
The DM-electron cross section above $10^{-29}(10^{-28})~{\rm cm}^2$ is excluded for $m_\chi\approx 0.1(10)$ MeV.

\begin{figure}[htb!]
\begin{center}
\includegraphics[scale=1,width=0.8\linewidth]{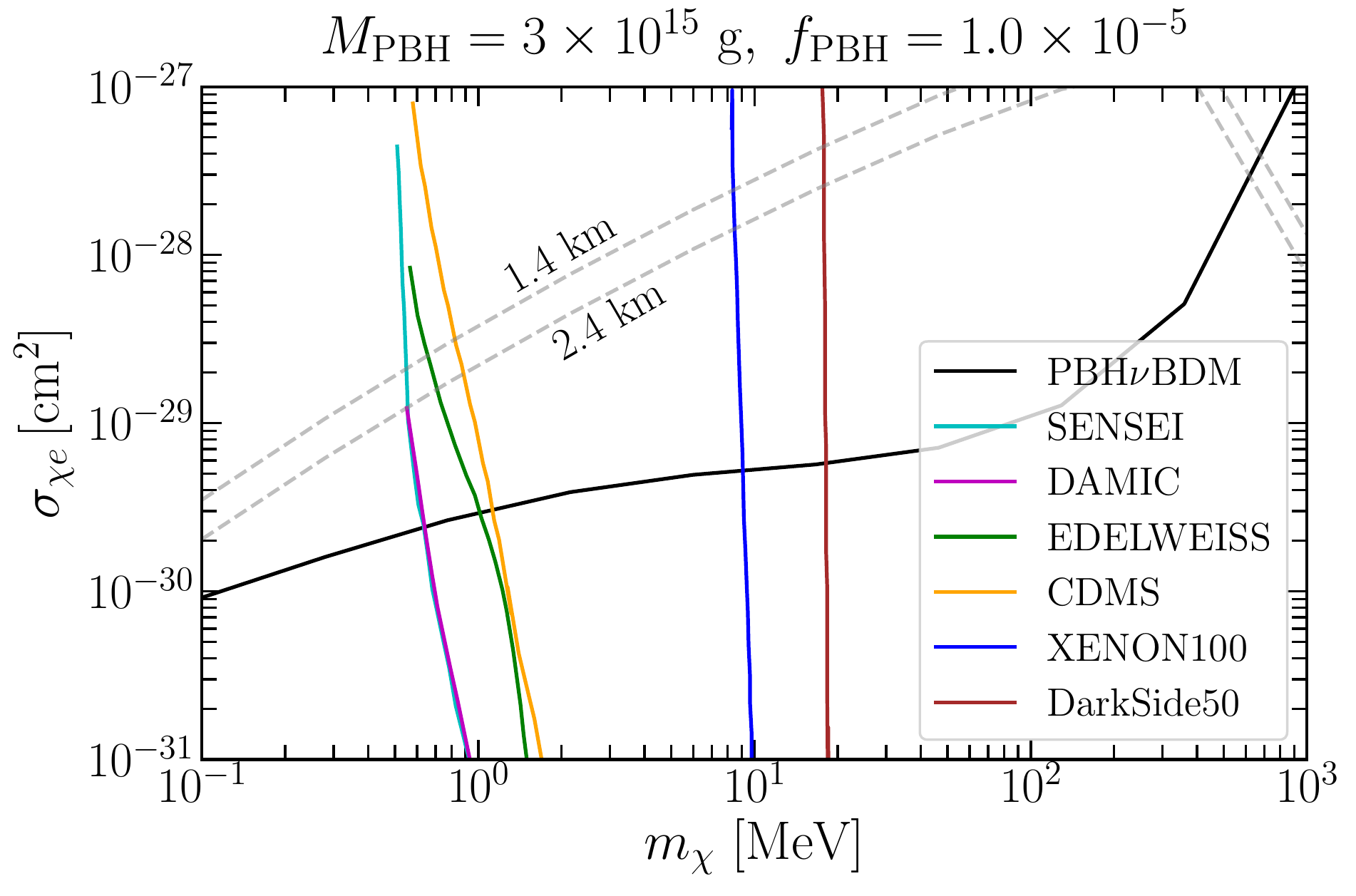}\\
\includegraphics[scale=1,width=0.8\linewidth]{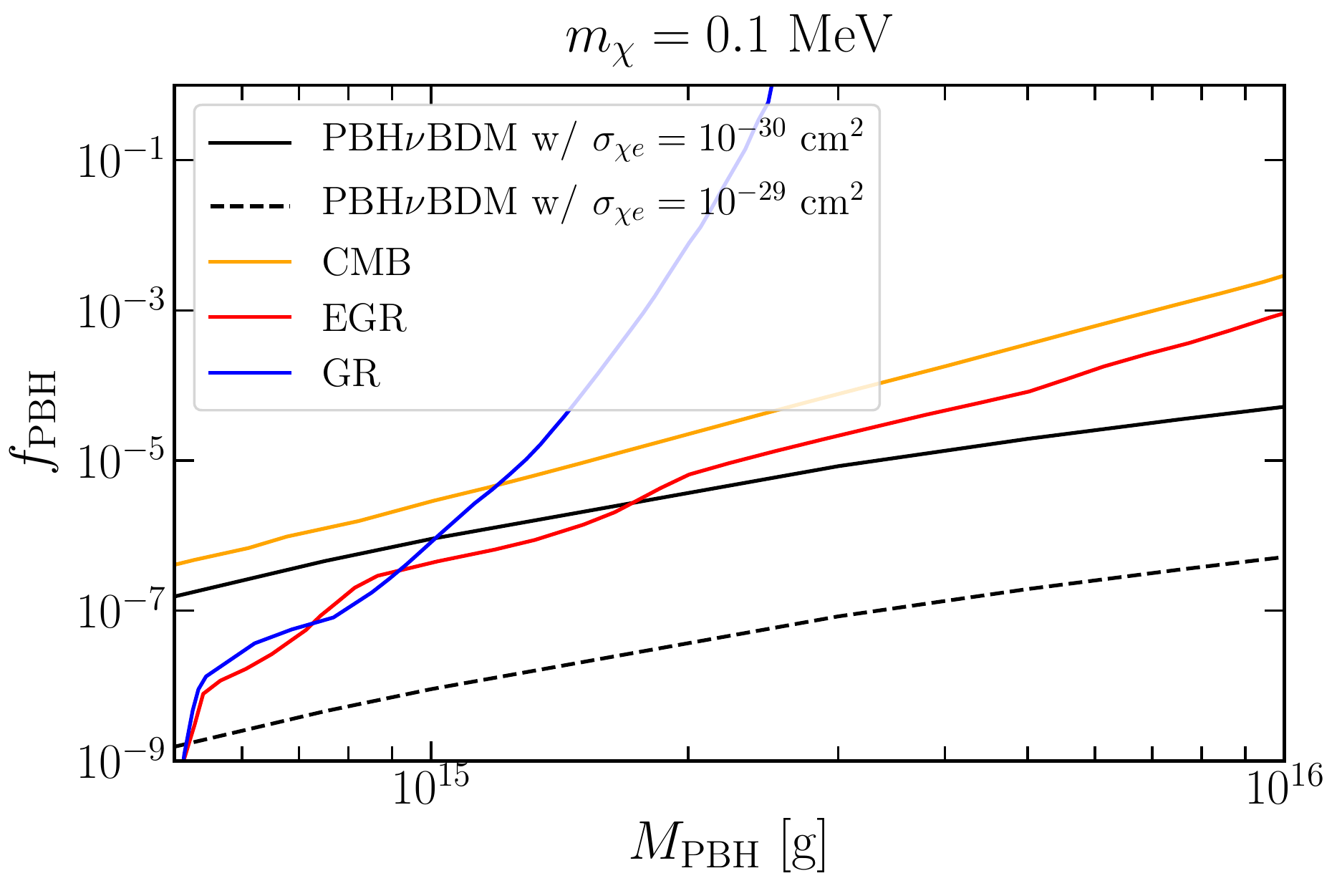}
\end{center}
\caption{Top: Projected upper bound on $\sigma_{\chi e}$ as a function of $m_\chi$ for XENON experiment. We assume $M_{\rm PBH}=3\times 10^{15}$ g, $f_{\rm PBH}=10^{-5}$. The constraints from DM direct detection experiments are also shown, including SENSEI~\cite{SENSEI:2020dpa} (cyan), DAMIC~\cite{DAMIC:2019dcn} (magenta), EDELWEISS~\cite{EDELWEISS:2020fxc} (green), CDMS~\cite{SuperCDMS:2018mne} (orange), XENON100~\cite{Essig:2017kqs} (blue) and DarkSide50~\cite{DarkSide:2018ppu} (brown). The Earth shielding limit is denoted by dashed gray curve for DM particles traveling a distance of 1.4 km for LNGS or 2.4 km for CJPL. Bottom: Projected upper bound on $f_{\rm PBH}$ as a function of $M_{\rm PBH}$. We assume $m_\chi=0.1$ MeV, and $\sigma_{\chi e}=10^{-30}~{\rm cm}^2$ (black solid) or $\sigma_{\chi e}=10^{-29}~{\rm cm}^2$ (black dashed). The evaporation constraints in Ref.~\cite{Carr:2020gox} are also shown, including CMB~\cite{Acharya:2020jbv,Chluba:2020oip} (orange), extragalactic gamma-rays~\cite{Carr:2009jm} (red) and galactic gamma-rays~\cite{Carr:2016hva} (blue).
}
\label{fig:limit}
\end{figure}

Next we can estimate the projected upper bound on $\sigma_{\chi e}$ as a function of $m_\chi$ for future XENON experiments. We impose a hard recoil cutoff at 4 keV and calculate the projected bounds by requiring 5 or more events over an exposure of one ton year.
This can be converted into the upper bound on $\sigma_{\chi e}$ as shown in the top panel of Fig.~\ref{fig:limit}. For $m_\chi=0.1$ MeV, the bound reaches down to $10^{-30}~{\rm cm}^2$ and can be further improved by at least one order of magnitude for the planned LZ~\cite{LZ:2019sgr}, XENONnT~\cite{XENON:2020kmp} and DARWIN~\cite{DARWIN:2016hyl} experiments with more than 20 ton year exposure. The constraints from DM direct detection experiments are also added and the regions to the right of the curves have been excluded. Following Refs.~\cite{Starkman:1990nj,Das:2021lcr}, the Earth shielding limit is denoted by dashed gray curve for DM particles traveling a distance of 1.4 km for the Gran Sasso Underground Laboratory (LNGS) or 2.4 km for the China Jinping Underground Laboratory (CJPL). One can see that the future experiments have potential to probe the PBH$\nu$BDM scenario for $m_\chi\lesssim 0.5$ MeV. This bound can also be translated into the upper limit on $f_{\rm PBH}$ as a function of $M_{\rm PBH}$ as seen in the bottom panel of Fig.~\ref{fig:limit}, for different DM-electron cross sections and $m_\chi=0.1$ MeV. For $\sigma_{\chi e}=10^{-30}~{\rm cm}^2$, compared with the current evaporation constraint from extragalactic gamma-rays, an improvement of $f_{\rm PBH}$ bound can be achievable for $M_{\rm PBH}\gtrsim 2\times 10^{15}$ g.

\section{Conclusions}
\label{sec:Con}

The primordial black holes and the boosted sub-GeV DM particle in the Milky Way are two interesting DM candidates beyond the WIMP. We explore the boosted DM by the neutrino flux from PBH evaporation to connect the macroscopic PBHs to sub-GeV DM particle. We consider this PBH$\nu$BDM scenario to interpret the XENON1T keV excess and find the XENON1T data together with the background model favor $m_\chi\sim 300$ MeV and $\sigma_{\chi e}\sim 10^{-28}~{\rm cm}^2$. Using the events predicted by the PBH$\nu$BDM scenario only, the DM-electron cross section above $10^{-29}(10^{-28})~{\rm cm}^2$ is excluded for $m_\chi\approx 0.1(10)$ MeV at 95\% CL. We also impose the projected bounds on the sub-GeV DM-electron scattering cross section and the fraction of DM composed of PBHs $f_{\rm PBH}$ in future XENON experiments. Given an exposure of one ton year, the future experiments have potential to probe the PBH$\nu$BDM scenario for $m_\chi\lesssim 0.5$ MeV.
Compared with the current evaporation constraint from extragalactic gamma-rays, the bound of $f_{\rm PBH}$ can be substantially improved.

Note: There is also a possibility that PBHs directly emit light DM particles in a very recent literature~\cite{Calabrese:2021src}.
We find that their boosted DM flux is comparable with our PBH$\nu$BDM flux shown in Fig.~\ref{fig:BDMflux}. Unlike us, they studied the BDM-nucleus scattering.


\acknowledgments
TL would like to thank Xin-He Meng for useful discussions.
TL is supported by the National Natural Science Foundation of China (Grant No. 11975129, 12035008) and ``the Fundamental Research Funds for the Central Universities'', Nankai University (Grant No. 63196013). JL is supported by the National Natural Science Foundation of China (Grant No. 11905299), Guangdong Basic and Applied Basic Research Foundation (Grant No. 2020A1515011479), the  Fundamental  Research  Funds  for  the  Central Universities, and the Sun Yat-Sen University Science Foundation. WC is supported by the National Natural Science Foundation of China under grant No. 11775025 and the Fundamental Research Funds for the Central Universities under grant No. 2017NT17.

\appendix

\bibliography{refs}

\end{document}